\documentclass[cits]{pos}

\newcommand{\be}{\begin{equation}}
\newcommand{\ee}{\end{equation}}
\newcommand{\ba}{\begin{eqnarray}}
\newcommand{\ea}{\end{eqnarray}}
\newcommand{\chpt}{$\chi$PT}
\newcommand{\cO}{{\cal O}}
\newcommand{\cJ}{{\cal J}}

\newcommand{\Leff}{{L_{10}^{\rm eff}}}
\newcommand{\Ceff}{{C_{87}^{\rm eff}}}
\newcommand{\Lmu}{{L_{10}^r(\mu)}}

\newcommand{\Lrho}{{L_{10}^r(M_\rho)}}
\newcommand{\Lninerho}{{L_{9}^r(M_\rho)}}
\newcommand{\Crho}{{C_{87}^r(M_\rho)}}

\title{Chiral low-energy constants from tau data}

\ShortTitle{Chiral low-energy constants from tau data}
\author{\speaker{Mart\'in Gonz\'alez-Alonso}%
	\\ Departament de F\'{\i}sica Te\`orica, IFIC, Universitat de Val\`encia-CSIC,\\
	Apt. Correus 22085, E-46071 Val\`encia, Spain\\
	E-mail: \email{Martin.Gonzalez@ific.uv.es}}
\author{Antonio Pich
	\\ Departament de F\'{\i}sica Te\`orica, IFIC, Universitat de Val\`encia-CSIC,\\
	Apt. Correus 22085, E-46071 Val\`encia, Spain\\
	E-mail: \email{Antonio.Pich@ific.uv.es}}
\author{Joaquim Prades
	\\ CAFPE and Departamento de F\'{\i}sica Te\'orica y del Cosmos,\\
	Universidad de Granada, Campus de Fuente Nueva, E-18002 Granada, Spain\\
	E-mail: \email{Prades@ugr.es}}

\abstract{We analyze how the recent precise hadronic $\tau$-decay data on the $V-A$ spectral function and general properties of QCD such as analyticity, the operator product expansion and chiral perturbation theory (\chpt), can be used to improve the knowledge of some of the low-energy constants of \chpt. In particular we find the most precise values of $L_{9,10}^r$ (or equivalently $\overline l_{5,6}$) at order $p^4$ and $p^6$ and the first phenomenological determination of $C_{87}^r$ ($c_{50}^r$).}

\FullConference{6th International Workshop on Chiral Dynamics\\
		 July 6-10 2009\\
		 Bern, Switzerland}

\begin{document}


\section{Introduction}
The fact that the $\tau$ is the only lepton massive enough to decay into hadrons makes it an excellent tool to study QCD, both perturbative and non-perturbative, using the precise hadronic $\tau$-decay data provided by ALEPH \cite{ALEPH05}. The determination of the QCD coupling $\alpha_s(M_\tau)$ \cite{alphas,LDP:92,DHZ06,new08,review}, which becomes the most precise determination of $\alpha_s(M_Z)$ after QCD running, is an excellent example. In this particular case the non-perturbative contributions are strongly suppressed, but in other analyses the non-perturbative effects are sizable and then one can extract important phenomenological hadronic matrix elements and other non-perturbative QCD quantities. Thanks to the fact that the spectral function of the $\tau$ decay can be separated experimentally in its vector and axial-vector contributions, we can study their difference that is specially interesting because it vanishes in perturbative QCD (in the chiral limit) and therefore it is a purely non-perturbative quantity.

The $\tau$-decay measurement of this $V-A$ spectral function has been used to perform \cite{DG:94,DHG98,NAR01} phenomenological tests of the so-called Weinberg sum rules (WSRs) \cite{WSR}, to compute the electromagnetic mass difference of the pions \cite{DHG98}, and to determine several QCD vacuum condensates \cite{DS07,CGM03} relevant for the computation of $\varepsilon_K'/\varepsilon_K$ \cite{Q7Q8}.
The common idea under these studies is the use of the analyticity properties of the different two-point correlation functions appearing in
the dynamical description of the $\tau$ hadronic width. As it is well known, analyticity 
allows us to relate different regions of the $q^2$-complex plane. Roughly speaking, one can relate in this way regions where we are able to compute analytically, either with Chiral Perturbation Theory (\chpt ) or with the short-distance Operator Product Expansion (OPE), with regions 
where we are not able to compute (except perhaps in the lattice) but that are experimentally accessible. This connection can be used either to predict observables that we are not able to calculate ``directly'' or, in the other way around, to extract the value of QCD parameters that are not fixed theoretically.

Using \chpt\ \cite{WEI79,GL84,GL85}, the hadronic $\tau$-decay data can also be related to order parameters of the spontaneous chiral symmetry breaking (S$\chi$SB) of QCD. \chpt\ is the effective field theory of QCD at very low energies that describes the physics of the S$\chi$SB Nambu-Goldstone bosons through an expansion in external momenta and quark masses, with coefficients that are order parameters of S$\chi$SB. At lowest order (LO), i.e. $\cO(p^2)$, all low-energy observables are described in terms of the pion decay constant $f_\pi \simeq 92.4$ MeV and the light quark condensate. At $\cO(p^4)$, the SU(3) \chpt\ Lagrangian contains 12 low-energy constants (LECs), $L_{i=1,\cdots,10}$ and $H_{1,2}$ \cite{GL85}, whereas at $\cO(p^6)$ we have 94  (23) additional parameters $C_{i=1,\cdots,94} \; (C^W_{i=1,\cdots,23})$ in the even (odd) intrinsic parity sector \cite{p6}. These LECs are not fixed by symmetry requirements alone and have to be determined phenomenologically or using non-perturbative techniques. Values for the $L_i$ couplings have been obtained in the past with an acceptable accuracy (a recent compilation can be found in ref.~\cite{ECK07}), but much less well determined are the ${\cal O}(p^6)$ couplings $C_i$.

There has been a lot of recent activity to determine the chiral LECs analytically, using as much as possible QCD information \cite{MOU97,KN01,RPP03,BGL03,CEE04,CEE05,KM06,RSP07,MP08,PRS08}, and from lattice simulations \cite{Shintani:2008qe,Boyle:2009xi,Frezzotti:2008dr,Aoki:2009qn}. This strong effort is motivated by the precision required in present phenomenological applications, which makes necessary to include corrections of $\cO(p^6)$ where the huge number of unknown couplings is the major source of theoretical uncertainty.

Here we explain how the determination of some of these LECs can be improved significantly using the most recent experimental data on hadronic $\tau$ decays \cite{ALEPH05}. In particular we will obtain the most accurate results for the \chpt\ couplings $L_{9}$, $L_{10}$ and $C_{87}$ or equivalently, in the $SU(2)$ \chpt\, language, $\overline l_5$, $\overline l_6$
and $c_{50}$ \cite{GPP08}.
Previous work on $L_{10}$ using $\tau$-decay data can be found in refs.~\cite{DHG98,NAR01,DS07,DS04}. Our analysis is the first one which includes the known two-loop \chpt\ contributions and, therefore, provides also the
SU(3) (SU(2)) $\cO(p^6)$ couplings $C_{87}$ ($c_{50}$).

We will first introduce the sum rule relations that we will use, then we will show our results and finally we will compare them with other recent analytic results and hadronic $\tau$-data determinations.


\section{Sum rule approach}
\begin{figure}[hbt]
\centering
\includegraphics[width=0.35\textwidth]{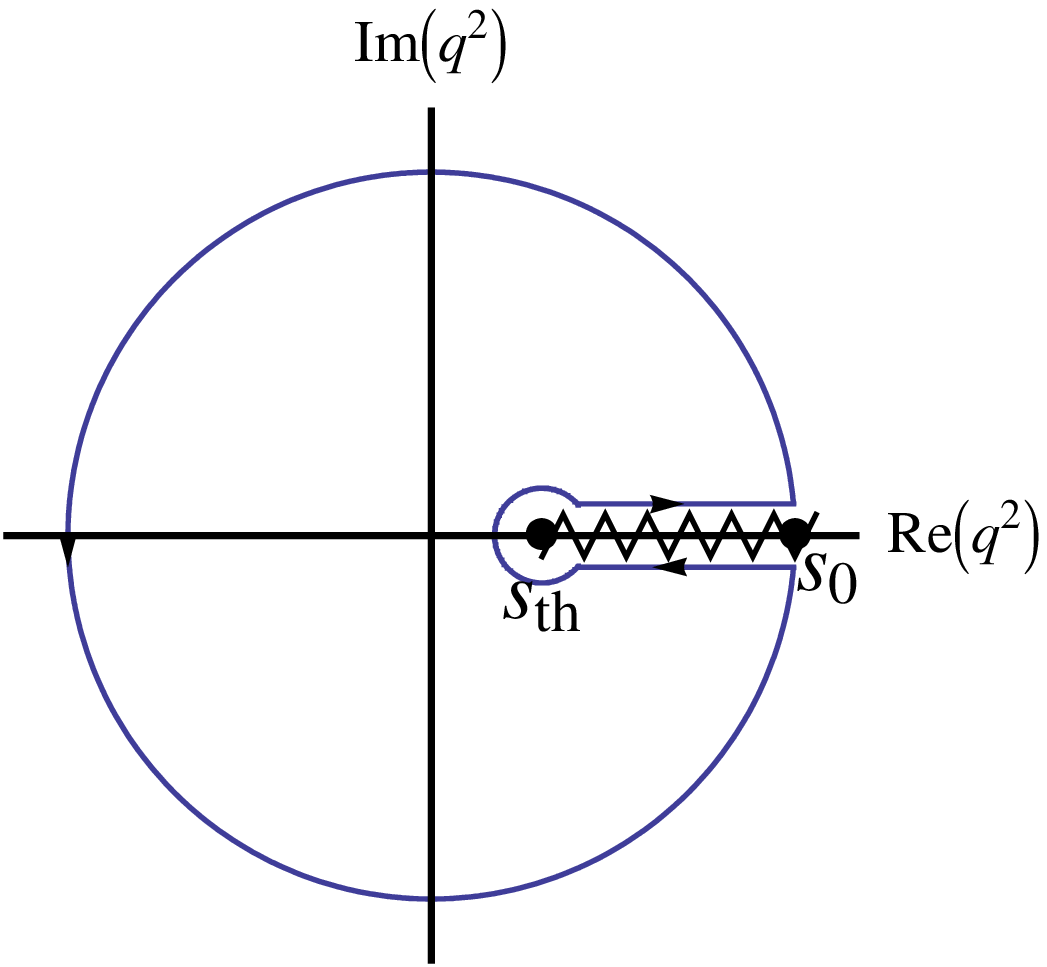}
\caption[]{Analytic structure of $\overline{\Pi}(s)$.}
\label{fig:circuit}
\end{figure}
%
The basic objects of the theoretical analysis are the two-point correlation functions of the non-strange vector 
($\cJ_{ud}^\mu = V_{ud}^\mu = \overline{u}\, \gamma^\mu d$)
and axial-vector ($\cJ_{ud}^\mu = A_{ud}^\mu = \overline{u}\, \gamma^\mu \gamma_5 d$) quark currents:
\ba
\label{eq:two}
\Pi^{\mu\nu}_{ud,\cJ}(q) &\;\equiv\; & i  \int \mathrm{d}^4 x \;  \mathrm{e}^{i q x}\;
\langle 0 | T \!\left( \cJ_{ud}^\mu(x) \cJ_{ud}^\nu(0)^\dagger \right)\! | 0 \rangle \nonumber \\
&\; =\; & (-g^{\mu\nu} q^2 + q^\mu q^\nu )\; \Pi^{(1)}_{ud,\cJ}(q^2)\; +\; q^\mu q^\nu\; \Pi^{(0)}_{ud,\cJ}(q^2)\, .
\ea
%
In particular, we are interested in the difference $\Pi(s) \equiv \Pi_{ud,V}^{(0+1)}-\Pi_{ud,A}^{(0+1)}$, and we will work in the isospin limit ($m_u=m_d$) where $\Pi^{(0)}_{ud,V}(q^2)=0$. The analytic behaviour of this correlator is shown in Fig.\ref{fig:circuit}, together with the complex circuit that we will use to apply Cauchy's theorem.
As we are interested in relating the \chpt\ domain (very low energies) with the $\tau$ data, we multiply this correlator by a weight function of the form $1/s^n$ with $n>0$. In this way we generate a residue at $s=0$.
Taking into account the OPE associated with our correlator at large momenta and working 
with the cases $n=1,2$, one gets the following sum rules (see ref.~\cite{GPP08} for a careful derivation):
\ba
\label{eq:SR1}
-8 \, L_{10}^{\rm eff} &\;\equiv\; & \int^{\infty}_{s_{\rm th}} \frac{\mathrm{d}s}{s}\; \frac{1}{\pi} \, {\rm Im} \, \Pi(s)
\; =\; \frac{2 f_\pi^2}{m_\pi^{2}}\, +\, \Pi(0)~, 
\\
\label{eq:SR2}
16 \, C_{87}^{\rm eff} &\;\equiv\; & \int^{\infty}_{s_{\rm th}} \frac{\mathrm{d}s}{s^2}\; \frac{1}{\pi} \, {\rm Im} \, \Pi(s)
\; =\; \frac{2 f_\pi^2}{m_\pi^{4}}\, +\, \frac{\rm{d}\Pi}{\rm{ds}}(0)~,
\ea
where the integrations start at the threshold $s_{\rm th}=4 m_\pi^2$.
These two relations represent the starting point of our work and define the effective parameters $\Leff$ and $\Ceff$. Their interest stems from the fact that the l.h.s. can be extracted from the data (see Section \ref{sec:eff}), while the r.h.s. can be rigourously calculated within \chpt~in terms of the LECs that we want to determine (see Section \ref{sec:real}).

\section{Determination of the effective parameters}
\label{sec:eff}
\begin{figure}[thb]
\vfill
\centerline{\begin{minipage}[c]{.47\linewidth}
\centering
\centerline{\includegraphics[width=7cm]{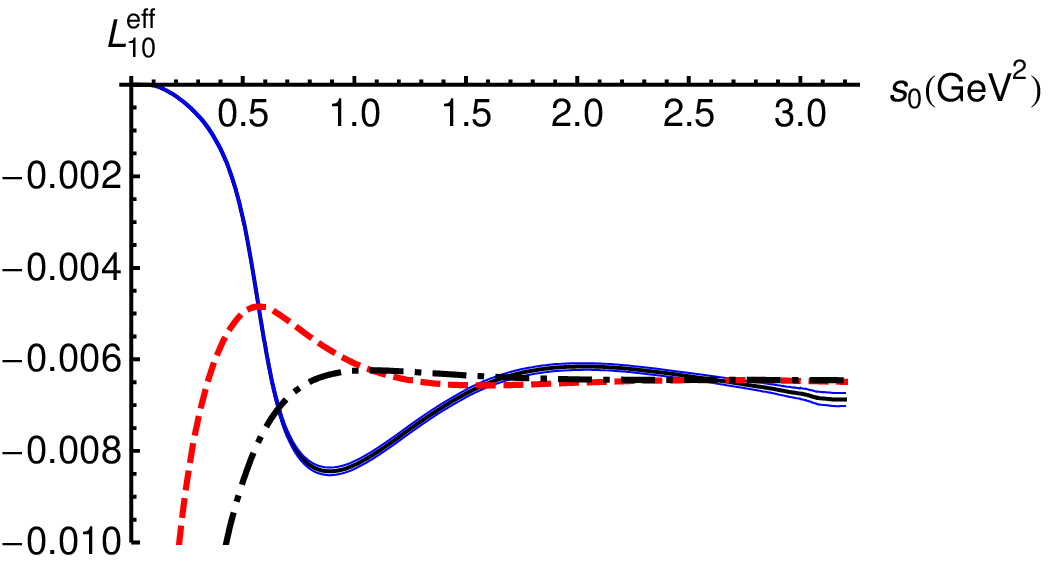}}
\end{minipage}
\hspace{0.67cm}
\begin{minipage}[c]{.47\linewidth}\centering
\centerline{\includegraphics[width=7cm]{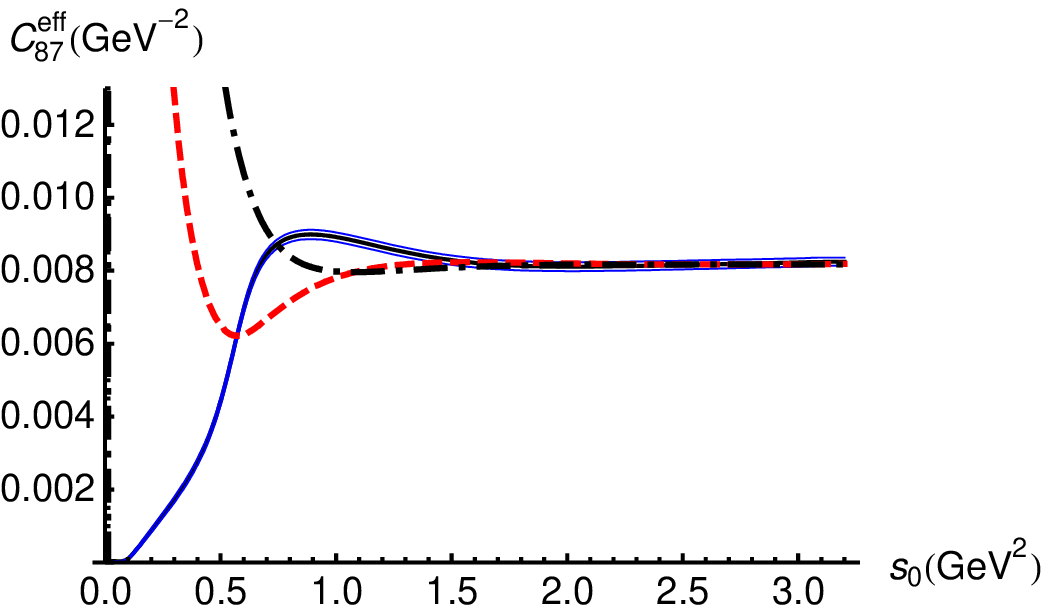}}
\end{minipage}
}
\vfill
\caption{$L_{10}^{\rm eff}(s_0)$ and $C_{87}^{\rm eff}(s_0)$ from different sum rules. For clarity, we do not include the error bands associated with the modified weights.}
\label{fig:L10C87}
\end{figure}

We use the recent ALEPH data on hadronic $\tau$ decays \cite{ALEPH05}, that provide the most precise measurement of the $V-A$ spectral function. In the integrals of equations (\ref{eq:SR1}) and (\ref{eq:SR2}) we are forced to cut the integration at a finite value $s_0$, neglecting in this way the rest of the integral from $s_0$ to infinity. The superconvergence properties of $\Pi(s)$ at large momenta imply a tiny contribution from the neglected range of integration, provided $s_0$ is large enough. Nevertheless, this
generates a theoretical error called quark-hadron duality violation (DV)\footnote{From a different but equivalent perspective, we are assuming that the OPE is a good approximation for $\Pi(s)$ at any $|s|\!\!=\!\!s_0$, what is not expected to happen near the real axis, and that produces the DV.}. From the $s_0$-sensitivity of the effective parameters one can assess the size of this error.

In Fig. \ref{fig:L10C87}, we plot the value of $L_{10}^{\rm eff}$ obtained for different values of $s_0$ (solid lines), with the one-sigma experimental error band, and we can see a quite stable result at $s_0\!\gtrsim\! 2~\mathrm{GeV}^2$. The weight function $1/s$ decreases the impact of the high-energy region, minimising the DV. The resulting integral is then much better behaved than the corresponding sum rules with $s^n$ ($n\ge0$) weights.

There are some possible strategies to estimate the value of $L_{10}^{\rm eff}$ and his error. One is to give the predictions fixing $s_0$ at the so-called ``duality points'', two points where the first and second 
WSRs \cite{WSR} happen to be satisfied. In this way we get $L_{10}^{\rm eff} = -(6.50\pm 0.13) \cdot 10^{-3}$, where the uncertainty covers the values obtained at the two ``duality points''. If we assume that the integral (\ref{eq:SR1}) oscillates around his asymptotic value with decreasing oscillations and we perform an average between the maxima and minima of the oscillations we get $L_{10}^{\rm eff} = -(6.5\pm 0.2) \cdot 10^{-3}$. Another way of estimating the DV uses appropriate oscillating functions defined in \cite{GON07} which mimic the real quark-hadron oscillations above the data. These functions are defined such that they match the data at $\sim\!3~\mathrm{GeV}^2$, go to zero with decreasing oscillations and satisfy the two WSRs. We find in this way $L_{10}^{\rm eff} = -(6.50\pm 0.12) \cdot 10^{-3}$, where the error spans the range generated by the different functions used. These estimates are in good agreement with each other and give us a first determination, but the most precise way to evaluate the error can be obtained taking advantage of the WSRs to construct modified sum rules with weight factors $w(s)$ proportional to $(1-s/s_0)$, in order to suppress numerically the role of the suspect region around $s\!\sim\!s_0$ \cite{LDP:92}. Fig.~\ref{fig:L10C87} shows the results obtained with $w_1(s)\!\equiv\!\left(1\!-\!s/s_0\right)/s$ (dashed line) and $w_2(s)\!\equiv\!\left(1\!-\!s/s_0\right)^2/s$ (dot-dashed line). These weights give rise to very stable results over a quite wide range of $s_0$ values. One gets $L_{10}^{\rm eff} = -(6.51\pm 0.06) \cdot 10^{-3}$ using $w_1(s)$ and $L_{10}^{\rm eff} = -(6.45\pm 0.06) \cdot 10^{-3}$ using $w_2(s)$. Taking into account all the previous discussion, we quote as our final result:
\be
\label{L10eff}
L_{10}^{\rm eff} = -(6.48\pm 0.06) \cdot 10^{-3} \, .
\ee

We have made a completely analogous analysis to determine $C_{87}^{\rm eff}$. The results are shown in Fig.~\ref{fig:L10C87}. The solid lines, obtained from Eq.~(\ref{eq:SR2}), are much more stable than the corresponding results for $L_{10}^{\rm eff}$, due to the $1/s^2$ factor in the integrand. The dashed and dot-dashed lines have been obtained with the modified weights $\rm{w_3(s)\equiv\! \frac{1}{s^2}\!\left(1\!-\!\frac{s^2}{s_0^2}\right)}$ and $\rm{w_4(s)\equiv\! \frac{1}{s^2} \left(1\!-\!\frac{s}{s_0}\right)^2 \left(1\!+\!2\frac{s}{s_0}\right)}$. The agreement among the different estimates is quite remarkable, and our final result is
\be
\label{C87eff}
C_{87}^{\rm eff} = (8.18\pm 0.14) \cdot 10^{-3} \, {\rm GeV}^{-2} \, .
\ee
Our result for $L_{10}^{\rm eff}$ agrees with \cite{DHG98,DS07,DS04}, but our estimation includes a more careful assessment of the theoretical errors. The $\rm{3.2\,\sigma}$ discrepancy between the estimation of ref.~\cite{NAR01} and ours is caused by an underestimation of the systematic error associated with the duality-point approach used in that reference. Only in ref.~\cite{DS04} $C_{87}^{\rm eff}$ is also determined and it is in good agreement with our result.

\section{Determination of the \chpt\ couplings}
\label{sec:real}
Using the results of ref.~\cite{ABT00} to calculate in \chpt\ the r.h.s. of equations (\ref{eq:SR1}) and (\ref{eq:SR2}), we get
\ba
\label{L10-p6}
-8~\Leff &\; =&\; -~8~L_{10}^r(\mu) + G^4_{1L}(\mu)  +~G^6_{0L}(\mu) + G^6_{1L}(\mu) + G^6_{2L}(\mu) ~, \\
\label{C87-p6}
16~\Ceff &\; =&\; H^4_{1L} +~16~C_{87}^r(\mu)+H^6_{1L}(\mu)+H^6_{2L}(\mu) ~,
\ea
where the functions $G^m_{nL}(\mu), H^{m}_{nL}(\mu)$ are corrections of order $p^m$ generated at the $n$-loop level, which explicit analytic form \cite{GPP08} is omitted for simplicity.

Working at $\cO(p^4)$, the determination of the chiral coupling $L_{10}$ is straightforward. One gets
\be
\label{valL10p4}
L_{10}^r(\mu\!=\!M_\rho) = -(5.22 \pm 0.06) \cdot 10^{-3}  \, .
\ee
At order $p^6$, the numerical relation is more involved because the small corrections $G^6_{0L,1L}(\mu)$ contain some LECs that represent the main source of uncertainty for $L_{10}^r$. It is useful to classify the $\cO(p^6)$ contributions through their ordering within the $1/N_C$ expansion. The tree-level term $G_{0L}^6(\mu)$ contains the only $\cO(p^6)$ correction in the large--$N_C$ limit, $\rm{4 m_\pi^2 (C_{61}^r \!-\! C_{12}^r \!-\! C_{80}^r)}$; this correction is numerically small because of the $m_\pi^2$ suppression and can be estimated with a moderate accuracy \cite{CEE05,KM06,ABT00,JOP04,UP08}. At NLO $G_{0L}^6(\mu)$ contributes with a term of the form $m_K^2(C_{62}^r \!-\! C_{13}^r \!-\! C_{81}^r)$. In the absence of information about these LECs we will adopt the conservative range $|C_{62}^r \!- C_{13}^r \!- C_{81}^r\!| \le |C_{61}^r \!- C_{12}^r \!- C_{80}^r|/3$, which generates the uncertainty that will dominate our final error on $L_{10}^r$. Also at this order in $1/N_C$ there is the one-loop correction $G_{1L}^6(\mu)$, that is proportional to $L_{9}^r$ which is better known \cite{BT02}.
Calculating the $1/N_C^2$ suppressed two-loop function $G_{2L}^6(\mu)$ and taking all these contributions into account we finally get the wanted $\cO(p^6)$ result:
\ba
\label{valL10p6}
\Lrho &\; =\; & -(4.06 \pm 0.04_{L_{10}^{\mathrm{eff}}}\pm 0.39_{\mathrm{LECs}}) \cdot 10^{-3} 
\; =\; -(4.06 \pm 0.39) \cdot 10^{-3} \, ,
\ea
where the error has been split into its two main components.

A recent reanalysis of the decay $\pi^+ \to e^+ \nu \gamma$  \cite{UP08}, using new experimental data, has provided quite accurate values for the combination $L_9+L_{10}$, both at order $p^4$ and $p^6$, that combined with our results for $\Lrho$ give us
\be
\Lninerho \, =\,\left\{
\begin{array}{ccc}
 (6.54 \pm 0.15) \cdot 10^{-3}  & \qquad\quad & [\cO(p^4)],\\[7pt]
 (5.50 \pm 0.40) \cdot 10^{-3}  & \qquad & [\cO(p^6)].
\end{array}\right.
\ee
Repeating the process we have done for $L_{10}^r$ with $C_{87}^r$ (where the only LEC involved is $L_9^r$) we get
\be
\label{valC87}
\Crho = (4.89 \pm 0.19) \cdot 10^{-3} \:\mathrm{GeV}^{-2}\, .
\ee

\section{SU(2)  \chpt}
Up to now, we have discussed the LECs of the usual SU(3) \chpt\ (\chpt$_3$). It turns useful to consider also the effective low-energy theory with only two flavours of light quarks (\chpt$_2$). In some cases, this allows to perform high-accuracy phenomenological determinations of the corresponding LECs at  NLO. Moreover, recent lattice simulations \cite{Aoki:2009qn,Frezzotti:2008dr} with two dynamical quarks are already able to obtain the SU(2) LECs with sufficient accuracy and this is an important check for them.

In SU(2) \chpt, there are ten LECs, $l_{i=1,..7}$ and $h_{1,2,3}$, at $\cO(p^4)$ (NLO) \cite{GL84}. Using the $\cO(p^6)$ relation between $l_5^r(\mu)$  and $\Lmu$, recently obtained in ref.~\cite{GHI}, we get
\be
\overline l_5 \, =\,\left\{
\begin{array}{ccc}
 13.30 \pm 0.11 & \qquad\quad & [\cO(p^4)],\\[7pt]
 12.24 \pm 0.21 & \; & [\cO(p^6)].
\end{array}\right.
\ee
Analogously to the $SU(3)$ case, the combination $\overline l_6 - \overline l_5$ has been determined from the analysis of $\pi \to l \nu \gamma$ \cite{BT97}. In combination with our determinations for $\overline l_5$ this gives us\footnote{Actually, at order $p^4$, the most precise value of the combination $\overline l_6 - \overline l_5$ is obtained if we calculate it from the $SU(3)$ combination $L_9+L_{10}$ of ref. \cite{UP08}. In this way we have obtained a prediction for $\overline l_6$ that supersedes that of ref. \cite{GPP08}.}
\be
\overline l_6\, =\,\left\{
\begin{array}{ccc}
 15.80 \pm 0.29  & \qquad\quad & [\cO(p^4)],\\[7pt]
 15.22 \pm 0.39  & \; & [\cO(p^6)].
\end{array}\right.
\ee

Making use of the recent results obtained in reference \cite{GHI} we can also rewrite our result for $C_{87}^r$ in the \chpt$_2$ language, getting in this way the first determination of $c_{50}^r$
\be
c_{50}^r(M_\rho) = (4.95 \pm 0.19) \cdot 10^{-3}~\rm{GeV}^{-2}~.
\ee


\section{Summary and comparison with previous estimates}

Tables~\ref{tab:p6} and \ref{tab:p4} summarize our determinations of chiral LECs at $\cO(p^6)$ and $\cO(p^4)$, respectively.
They have been obtained through a sum rule analysis that only uses general properties of QCD and the measured $V\!-\!A$ spectral function \cite{ALEPH05}, and taking into account the results of refs. \cite{UP08,BT97}.

\begin{table}[tbh]\centering
\begin{tabular}{|c|c|}
	\hline
	\chpt$_2$ & \chpt$_3$ \\
	\hline
	$\overline l_5 = 12.24 \pm 0.21$	& $\Lrho = -(4.06\pm 0.39)\cdot 10^{-3}$  \\
	\hline
	$\overline l_6=  15.22 \pm 0.39$ 	& $\Lninerho= (5.50 \pm 0.40) \cdot 10^{-3}$  \\
	\hline
	$c_{50}^r=  (4.95 \pm 0.19) \cdot 10^{-3}~\rm{GeV}^{-2}$ 	& $\Crho = (4.89 \pm 0.19) \cdot 10^{-3}~\rm{GeV}^{-2}$  \\
	\hline
	\end{tabular}
\caption{Results for the \chpt\ LECs obtained at $\cO(p^6)$.}\label{tab:p6}
\end{table}

\begin{table}[tbh]\centering
	\begin{tabular}{|c|c|}
	\hline
	\chpt$_2$ & \chpt$_3$ \\
	\hline
	$\overline l_5 = 13.30 \pm 0.11$	& $\Lrho = -(5.22\pm 0.06)\cdot 10^{-3}$  \\
	\hline
	$\overline l_6=  15.80 \pm 0.29$ 	& $\Lninerho= (6.54 \pm 0.15) \cdot 10^{-3}$  \\
	\hline
	\end{tabular}
\caption{Results for the \chpt\ LECs obtained at $\cO(p^4)$.}\label{tab:p4}
\end{table}

Our determination of $L_{10}^r$ ($\overline l_5$) is the first one extracted from $\tau$-decay data at $\cO(p^6)$. We can make an indirect and interesting check comparing our $\cO(p^6)$ result for $L_9^r$ ($\overline l_6$) with the value $\Lninerho = (5.93 \pm 0.43) \cdot 10^{-3}$ \ ($\overline l_6= 16.0 \pm 0.5 \pm 0.7$) obtained from the charge radius of the pion \cite{BT02} (\cite{BCT98}). The agreement is very good and the improvement in the numerical value of $\overline l_6$ is remarkable.

At order $p^4$ we do have a previous estimate of $L_{10}^r$ from $\tau$ data \cite{DHG98} that found $\Lrho= - (5.13 \pm 0.19) \cdot 10^{-3}$, through a simultaneous fit of this parameter and the OPE corrections of dimensions six and eight to several spectral moments of the hadronic distribution. This determination is in good agreement with our $\cO(p^4)$ result. Our quoted uncertainty has an smaller experimental contribution and includes a better assessment of the theoretical uncertainties. We can also perform an indirect check through the comparison of our $\cO(p^4)$ result for $L_9^r$ with the value $\Lninerho = (6.9 \pm 0.7) \cdot 10^{-3}$ obtained from the charge radius of the pion \cite{ECK07}. We see again a very good agreement and a clear improvement in the precision.

If we shift now from phenomenology to theory, we can compare our results with those obtained from analytical approaches and lattice simulations. Our determinations of $\Lrho$ and $\Crho$ agree within errors with the large--$N_C$ estimates based on lowest-meson dominance \cite{KN01,CEE04,ABT00,PI02}, $L_{10} \approx -3 f_\pi^2 / (8 M_V^2) \approx -5.4\cdot 10^{-3}$ and $C_{87} \approx 7 f_\pi^2 / (32 M_V^4) \approx 5.3\cdot 10^{-3}~\mathrm{GeV}^{-2}$, and with the result of ref. \cite{MP08} for $C_{87}$, based on Pad\'e approximants. These predictions however are unable to fix the scale dependence which is of higher-order in $1/N_C$. More recently the resonance chiral theory Lagrangian \cite{CEE04,EGPdR89} has been used to analyse the correlator $\Pi(s)$ at NLO order in the $1/N_C$ expansion. Matching the effective field theory description with the short-distance QCD behaviour, both LECs are determined, keeping full control of the $\mu$-dependence. The predicted values $\Lrho = -(4.4 \pm 0.9) \cdot 10^{-3}$ and $\Crho=(3.6 \pm 1.3) \cdot 10^{-3}$  GeV$^{-2}$ \cite{PRS08} are in perfect agreement with our results, although less precise.

The most recent lattice calculations find the following results (order $p^4$):
\ba
\Lrho &=& \left\{
\begin{array}{lcl}
-(5.2 \pm 0.5) \cdot 10^{-3} &\qquad & \begin{minipage}{0.8cm}\cite{Shintani:2008qe}\end{minipage},\\
-(5.7 \pm 1.1 \pm 0.7) \cdot 10^{-3} &\qquad & \begin{minipage}{0.8cm}\cite{Boyle:2009xi}\end{minipage},
\end{array}\right.
\nonumber\\
\overline l_6&=&  \left\{
\begin{array}{ccc}
14.9 \pm 1.2\pm0.7 &\qquad &  \begin{minipage}{0.8cm}\cite{Frezzotti:2008dr}\end{minipage},\\
11.9 \pm 0.7\pm1.0 &\qquad & \begin{minipage}{0.8cm}\cite{Aoki:2009qn}\end{minipage}.
\end{array}\right.\ea
They are in good agreement with our determinations (although still far from the phenomenological precision), but for the last one that is slightly smaller. As discussed in ref.~\cite{Aoki:2009qn}, this is partly due to the deviation of the lattice determination of the pion decay constant from the \chpt~one.

Therefore we can conclude that the different analytical approaches and the various lattice calculations agree very well with our precise phenomenological values.
%
\section*{Acknowledgements}
M. G.-A. is indebted to MICINN (Spain) for an FPU Grant. Work partly supported by the EU network FLAVIAnet [MRTN-CT-2006-035482], by MICINN, Spain [FPA2007-60323, FPA2006-05294, CSD2007-00042 --CPAN--], by Junta de Andaluc\'{\i}a [P07-FQM 03048] and by Generalitat Valenciana [Prometeo/2008/069].

\end{document}